\documentclass[paper,notoc]{JHEP}
\usepackage{epsfig}

\def\gsim{\ \rlap{\raise 3pt \hbox{$>$}}{\lower 3pt \hbox{$\sim$}}\ }
\def\lsim{\ \rlap{\raise 3pt \hbox{$<$}}{\lower 3pt \hbox{$\sim$}}\ }
\title{Neutrino masses and mixings in non-factorizable geometry}

\author{Yuval Grossman\\
Stanford Linear Accelerator Center, Stanford University\\
Stanford, California 94309, U.S.A.\\
E-mail: \email{yuval@slac.stanford.edu}}

\author{Matthias Neubert\\
Newman Laboratory of Nuclear Studies, Cornell University\\
Ithaca, New York 14853, U.S.A.\\
E-mail: \email{neubert@mail.lns.cornell.edu}}

\abstract{We study bulk fermion fields in the localized gravity model 
with non-factorizable metric recently proposed by  Randall and Sundrum,
and Gogberashvili. In addition to a tower of weak-scale Kaluza--Klein 
states we find a zero mode for any value of the fundamental fermion 
mass. If the fermion mass is larger than half the curvature of the 
compact dimension, the zero mode can be localized on the ``hidden'' 
3-brane in the Randall--Sundrum model. Identifying this mode with a 
right-handed neutrino provides a new way for obtaining small Dirac 
neutrino masses without invoking a see-saw mechanism. Cancellation of 
the parity anomaly requires introducing an even number of bulk 
fermions. This naturally leads to a strong hierarchy of neutrino masses 
and generically large mixing angles.}


\preprint{CLNS~99/1656\\
SLAC-PUB-8330\\
December 1999\\[0.15cm]
\hepph{9912408}}

\begin{document}

\section{Introduction}

Theories with extra spatial dimensions have received great attention 
recently, when it was shown that they could provide a solution to the 
gauge-hierarchy problem. If space-time is a product of Minkowski 
space with $n$ compact dimensions, with Standard Model fields 
localized in the three extended spatial dimensions (i.e., on a 
3-brane) and gravity propagating in the extra space, then the 
strength of gravity on the 3-brane is governed by an effective Planck 
scale $M_{\rm Pl}^2=M^{n+2}\,V_n$, where $M$ is the fundamental scale 
of gravity and $V_n$ the volume of the compact space \cite{ADD}. If 
this space is sufficiently large, the fundamental scale $M$ can be of 
order 1\,TeV, thus removing the large disparity between the 
gravitational and the electroweak scales. 

An intriguing alternative to the above scenario invokes a 
non-factorizable geometry with a metric that depends on the 
coordinates of the extra dimensions \cite{Gogb,RS1}. In the simplest 
scenario due to Randall and Sundrum (RS) one considers a single extra 
dimension, taken to be a $S^1/Z_2$ orbifold parameterized by a 
coordinate $y=r_c\,\phi$, with $r_c$ the radius of the compact 
dimension, $-\pi\le\phi\le\pi$, and the points $(x,\phi)$ and 
$(x,-\phi)$ identified \cite{RS1}. There are two 3-branes located at 
the orbifold fixed points: a ``visible'' brane at $\phi=\pi$ 
containing the Standard Model fields, and a ``hidden'' brane at 
$\phi=0$. The solution of Einstein's equations for this geometry leads 
to the non-factorizable metric
\begin{equation}\label{metric}
   \mbox{d}s^2 = e^{-2k r_c|\phi|}\,\eta_{\mu\nu}\,
   \mbox{d}x^\mu \mbox{d}x^\nu - r_c^2\,\mbox{d}\phi^2 \,,
\end{equation}
where $x^\mu$ are the coordinates on the four-dimensional surfaces
of constant $\phi$, and the parameter $k$ is of order the fundamental
Planck scale $M$. (This solution can only be trusted if $k<M$, so the 
bulk curvature is small compared with the fundamental Planck scale.) 
The two 3-branes carry vacuum energies tuned such that 
$V_{\rm vis}=-V_{\rm hid}=-24M^3 k$, which is required to obtain a 
solution respecting four-dimensional Poincar\'e invariance. In between 
the two branes is a slice of AdS$_5$ space. 

With this setup, the effective Planck scale seen by particles 
confined to four-dimensional space-time is $M_{\rm Pl}^2
=(M^3/k)(1-e^{-2k r_c\pi})$, which is of order the fundamental scale 
$M$. Unlike the scenarios with large extra dimensions considered in 
\cite{ADD}, the scale $M$ is therefore not of order the weak scale. 
However, the ``warp factor'' $e^{-2k r_c|\phi|}$ in the metric 
(\ref{metric}) has important implications for the masses of particles 
confined to the visible brane. The Lagrangian for these particles 
depends on the induced metric on the brane, $g_{\mu\nu}^{\rm vis}
=e^{-2k r_c\pi}\,\eta_{\mu\nu}$, and after field renormalization any 
mass parameter $m_0$ in the fundamental theory is promoted into an 
effective mass parameter $m=e^{-k r_c\pi}\,m_0$ governing the physical 
properties of particles on the brane \cite{RS1}. With 
$k r_c\approx 12$ this mechanism produces weak-scale physical masses 
and couplings from fundamental masses and couplings of order the 
Planck scale. As a consequence of the warp factor, the Kaluza--Klein 
excitations of gravitons have weak-scale mass splittings and couplings 
\cite{RS2,LyRa}. This is in contrast with the Kaluza--Klein spectrum 
of gravitons propagating in large extra dimensions, which consists of 
a large number of light modes (with splittings of order the 
compactification scale) with gravitational couplings. The same 
properties (i.e., weak-scale masses and couplings) are shared by bulk 
scalars and vector particles propagating in the extra dimension 
\cite{GW1,DHR2,Poma}. 

In order for this model to provide a viable solution to the hierarchy 
problem it is important to address the question of how to stabilize 
the radius $r_c$ of the extra dimension, and the related question of 
the potentially disastrous cosmology of a visible universe confined 
to a brane with negative tension \cite{CGKT,CGS,SMS} (see also 
\cite{BDEL,FTW}). A mechanism for radius stabilization utilizing a 
bulk scalar field has been proposed in \cite{GW2}. In the presence of 
such a mechanism, one finds a conventional cosmological expansion for 
temperatures below the weak scale \cite{CGRT}. The couplings of the 
radion field to Standard Model particles may have interesting 
implications for collider searches \cite{CGRT,GW3}. Other 
phenomenological consequences of the RS model have been discussed in 
\cite{DHR1}, and an alternative model avoiding the negative-tension 
brane has been proposed in \cite{LyRa}.

Resolving the hierarchy problem by introducing extra dimensions poses 
new challenges. For instance, operators mediating proton decay, 
lepton-number violation or flavor-changing neutral currents must be 
sufficiently suppressed. Likewise, the see-saw mechanism for 
generating small neutrino masses cannot be invoked if the highest 
energy scale governing physics on the visible brane is the weak scale. 
However, there is now increasing evidence that the atmospheric 
neutrino anomaly \cite{AN1,AN2,AN3} and the solar neutrino problem 
\cite{SK,SN} are explained in terms of neutrino oscillations, which 
require small but non-vanishing neutrino masses. Several 
four-dimensional alternatives to the see-saw mechanism not requiring 
a high-energy scale have been proposed, such as radiatively generated 
neutrino masses \cite{Pal} and composite models \cite{AG}. However, 
it would be interesting to find new mechanisms that are intrinsically 
higher dimensional. In the context of models with large extra 
dimensions ideas in this direction have been presented in 
\cite{ADDM,ArSc}, and some concrete models have been worked out in 
\cite{DS,DaKo,Moha}. They contain a massless Standard Model singlet 
propagating in the bulk of the extra compact space, which serves as a 
right-handed neutrino. Then the effective four-dimensional Yukawa 
coupling is suppressed by a volume factor $1/\sqrt{V_n}$, reflecting 
the small overlap between the right-handed neutrino in the bulk and 
the left-handed one on the 3-brane. By construction, this factor 
provides a suppression of neutrino masses of order $v/M_{\rm Pl}$, 
reminiscent of the see-saw mechanism. However, this idea does not work 
in a scenario with small extra dimensions such as the RS model, simply 
because of the lack of a volume suppression factor. 

In this letter we investigate the possibility of incorporating bulk 
fermions in the RS model. As in the case of scalars \cite{GW1} and 
vector fields \cite{DHR2,Poma} propagating in the compact dimension 
we find that the Kaluza--Klein modes have weak-scale masses even 
though the fermion mass in the fundamental, five-dimensional theory 
is of order the Planck scale. The fermion case is more interesting, 
however, because the extension of the Dirac algebra to five dimensions 
leads to a different propagation of left- and right-handed modes. 
After imposing the orbifold boundary conditions the geometry supports 
a left-handed or a right-handed zero mode for any value of the 
fundamental fermion mass, one of which can be localized on the 
hidden brane of the RS model. This is different from the scalar and 
vector cases, where zero modes exist only for vanishing mass in the 
fundamental theory. The localization of a right-handed zero mode on 
the hidden brane provides a new mechanism for obtaining small 
neutrino masses, which can be realized by coupling the Higgs and 
left-handed lepton fields of the Standard Model, localized on the 
visible brane, to a right-handed fermion in the bulk. The neutrino 
mass can be tuned over many orders of magnitude by a small change of 
the bulk fermion mass. Moreover, cancellation of the parity anomaly 
\cite{Redl,Call} forces us to introduce an even number of bulk 
fermions. This naturally leads to a neutrino mass hierarchy and 
potentially large mixing angles.

\section{Bulk fermions}

Our starting point is the action for a Dirac fermion with mass $m$ 
of order the fundamental scale $M$ propagating in a five-dimensional 
space with the metric (\ref{metric}), which we write in the form 
\cite{book}\footnote{We do not include a five-dimensional Majorana 
mass term of the form $\Psi^T C\,\Psi$ in the action, because later 
we will assign lepton number to the bulk fermion.}
\begin{equation}\label{action}
   S = \int\!\mbox{d}^4x\!\int\!\mbox{d}\phi\,\sqrt{G}
   \left\{ E_a^A \left[ \frac{i}{2}\,\bar\Psi\gamma^a
   (\partial_A-\overleftarrow{\partial_A})\Psi
   + \frac{\omega_{bcA}}{8}\,\bar\Psi \{\gamma^a,\sigma^{bc}\} \Psi
   \right] - m\,\mbox{sgn}(\phi)\,\bar\Psi\Psi \right\} \,,
\end{equation}
where $G=\mbox{det}(G_{AB})=r_c^2\,e^{-8\sigma}$ with 
$\sigma=k r_c|\phi|$ is the determinant of the metric. We use capital
indices $A,B,\dots$ for objects defined in curved space, and 
lower-case indices $a,b,\dots$ for objects defined in the tangent 
frame. The matrices $\gamma^a=(\gamma^\mu,i\gamma_5)$ provide a 
four-dimensional representation of the Dirac matrices in 
five-dimensional flat space. The quantity
$E_a^A=\mbox{diag}(e^\sigma,e^\sigma,e^\sigma,e^\sigma,1/r_c)$ is the
inverse vielbein, and $\omega_{bcA}$ is the spin connection. Because 
in our case the metric is diagonal, the only non-vanishing entries of 
the spin connection have $b=A$ or $c=A$, giving no contribution to
the action in (\ref{action}).

The sign change of the mass term under $\phi\to-\phi$ is necessary in 
order to conserve $\phi$-parity, as required by the $Z_2$ orbifold 
symmetry of the RS model. Such a mass term can be obtained, e.g., by 
coupling the fermion to a pseudoscalar (under $\phi$-parity) bulk Higgs 
field. For a single bulk fermion in five dimensions $\phi$-parity is 
broken at the quantum level, giving rise to the so-called parity 
anomaly \cite{Redl,Call}. To cancel this anomaly, we will later 
consider an even number of fermion fields. 

Using an integration by parts, and defining left- and right-handed
spinors $\Psi_{L,R}\equiv\frac12(1\mp\gamma_5)\Psi$, the action can 
be written as
\begin{eqnarray}\label{Sdel}
   S &=& \int\!\mbox{d}^4x\!\int\!\mbox{d}\phi\,r_c\,\bigg\{
    e^{-3\sigma} \left( \bar\Psi_L\,i\rlap/\partial\,\Psi_L
    + \bar\Psi_R\,i\rlap/\partial\,\Psi_R \right) 
    - e^{-4\sigma}\,m\,\mbox{sgn}(\phi) \left( 
    \bar\Psi_L \Psi_R + \bar\Psi_R \Psi_L \right) \nonumber\\
   &&\mbox{}- \frac{1}{2 r_c} \left[ \bar\Psi_L \left(
    e^{-4\sigma} \partial_\phi + \partial_\phi\,e^{-4\sigma}
    \right) \Psi_R - \bar\Psi_R \left(
    e^{-4\sigma} \partial_\phi + \partial_\phi\,e^{-4\sigma}
    \right) \Psi_L \right] \bigg\} \,,
\end{eqnarray}
where we impose periodic boundary conditions $\Psi_{L,R}(x,\pi)
=\Psi_{L,R}(x,-\pi)$ on the fields. The action is even under the 
$Z_2$ orbifold symmetry if $\Psi_L(x,\phi)$ is an odd function of 
$\phi$ and $\Psi_R(x,\phi)$ is even, or vice versa. To perform the 
Kaluza--Klein decomposition we write
\begin{equation}\label{KK}
   \Psi_{L,R}(x,\phi) = \sum_n \psi_n^{L,R}(x)\,
   \frac{e^{2\sigma}}{\sqrt{r_c}}\,\hat f_n^{L,R}(\phi) \,.
\end{equation}
Because of the $Z_2$ symmetry of the action it is sufficient to 
restrict the integration over $\phi$ from 0 to $\pi$. The behavior of
the solutions for negative $\phi$ is then determined by their $Z_2$ 
parity. $\{ \hat f_n^L(\phi)\}$ and $\{ \hat f_n^R(\phi)\}$ are two 
complete, orthonormal (with a scalar product defined below) sets of 
functions on the interval $\phi\in[0,\pi]$, subject to certain 
boundary conditions. We will construct them as the eigenfunctions of 
hermitian operators on this interval. Inserting the ansatz (\ref{KK}) 
into the action and requiring that the result take the form of the 
usual Dirac action for massive fermions in four dimensions,
\begin{equation}\label{Sferm}
   S = \sum_n \int\!\mbox{d}^4x\,\Big\{
   \bar\psi_n(x)\,i\rlap/\partial\,\psi_n(x)
   - m_n\,\bar\psi_n(x)\,\psi_n(x) \Big\} \,,
\end{equation}
where $\psi\equiv\psi_L+\psi_R$ (except for possible chiral modes) 
and $m_n\ge 0$ , we find that the functions $\hat f_n^{L,R}(\phi)$ 
must obey the conditions
\begin{eqnarray}\label{cond}
   \int\limits_0^\pi\!\mbox{d}\phi\,e^\sigma
   \hat f_m^{L*}(\phi)\,\hat f_n^L(\phi)
   &=& \int\limits_0^\pi\!\mbox{d}\phi\,e^\sigma
    \hat f_m^{R*}(\phi)\,\hat f_n^R(\phi) = \delta_{mn} \,,
    \nonumber\\
   \left( \pm\frac{1}{r_c}\,\partial_\phi - m \right)
   \hat f_n^{L,R}(\phi) 
   &=& - m_n\,e^\sigma \hat f_n^{R,L}(\phi) \,.
\end{eqnarray}
The boundary conditions $\hat f_m^{L*}(0)\,\hat f_n^R(0)
=\hat f_m^{L*}(\pi)\,\hat f_n^R(\pi)=0$, which follow since either 
all left-handed or all right-handed functions are $Z_2$-odd, ensure
that the differential operators $(\pm\frac{1}{r_c}\,\partial_\phi-m)$ 
are hermitian and their eigenvalues $m_n$ real. (Since the equations 
are real, the functions $\hat f_n^{L,R}(\phi)$ could be chosen real 
without loss of generality.)

It is convenient to introduce the new variable $t=\epsilon\,e^\sigma
\in[\epsilon,1]$ with $\epsilon=e^{-k r_c\pi}$, rescale 
$\hat f_n^{L,R}(\phi)\to\sqrt{k r_c\epsilon}\,f_n^{L,R}(t)$, and 
define the quantities 
\begin{equation}
   \nu = \frac{m}{k} \,, \qquad
   x_n = \frac{m_n}{\epsilon k} = \frac{m_n}{k}\,e^{k r_c\pi} \,.
\end{equation} 
The small parameter $\epsilon\sim 10^{-16}$ sets the ratio between 
the electroweak and the gravitational scales. The two conditions in 
(\ref{cond}) now become
\begin{eqnarray}\label{cond1}
   \int\limits_\epsilon^1\!\mbox{d}t\,f_m^{L*}(t)\,f_n^L(t)
   &=& \int\limits_\epsilon^1\!\mbox{d}t\,f_m^{R*}(t)\,f_n^R(t)
    = \delta_{mn} \,, \nonumber\\
   (\pm t\,\partial_t-\nu) f_n^{L,R}(t)
   &=& - x_n t\,f_n^{R,L}(t) \,, 
\end{eqnarray}
and the boundary conditions are $f_m^{L*}(\epsilon)\,f_n^R(\epsilon)
=f_m^{L*}(1)\,f_n^R(1)=0$. The system of coupled, first-order 
differential equations for $f_n^{L,R}(t)$ implies the second-order
equations
\begin{equation}\label{eigen1}
   \left[ t^2 \partial_t^2 + x_n^2 t^2 - \nu(\nu\mp 1) \right]
   f_n^{L,R}(t) = 0 \,.
\end{equation}
Dimensional analysis shows that the eigenvalues $x_n$ are of order
unity, corresponding to weak-scale fermion masses $m_n$ in the 
four-dimensional theory.

The solution of the differential equations is straightforward. We 
start by looking for zero modes, i.e., solutions with $x_n=0$. In 
this case the first-order equations in (\ref{cond1}) decouple. The 
properly normalized solutions are
\begin{equation}\label{zero}
   f_0^{L,R}(t) = f_0^{L,R}(1)\,t^{\pm\nu} 
   \propto e^{\pm m r_c|\phi|} \,, \qquad
   |f_0^{L,R}(1)|^2 = \frac{1\pm 2\nu}{1-\epsilon^{1\pm 2\nu}} \,.
\end{equation}
Since these are even functions of $\phi$, which do not vanish at the
orbifold fixed points, only one of the zero modes is allowed by the 
orbifold symmetry. This mode exists irrespective of the value of the 
fermion mass $m$ in the five-dimensional theory. Note that for 
$\nu>\frac12$ the right-handed zero mode has a very small wave 
function on the visible brane: $f_0^R(1)\propto\epsilon^{\nu-\frac12}$. 
This property will allow us to obtain small neutrino masses. The 
presence of fermion zero modes should not come as a surprise, since it 
is well known that in flat space-time they are associated with domain 
walls \cite{Dom}. In our model the domain walls are provided by the 
3-branes of the RS model, which separate the regions with a different 
sign of the fermion mass term. The functions $P_L\,f_n^L(t)$ and 
$P_R\,f_n^R(t)$, with $P_{L,R}=\frac12(1\mp\gamma_5)$, can be 
associated with the ``fermionic'' and ``bosonic'' degrees of freedom 
of a supersymmetric, quantum-mechanical system \cite{KaSc}. The 
supersymmetry generators are $Q=(\partial_t-\nu/t)\gamma^0 P_L$ and 
$Q^\dagger=-(\partial_t+\nu/t)\gamma^0 P_R$, and the Kaluza--Klein
modes are the eigenstates of the Hamiltonian $\{Q,Q^\dagger\}$. This
explains why left- and right-handed modes have the same eigenvalues 
$x_n$. The two zero modes correspond to the ground-state solutions of 
the supersymmetric Hamiltonian. In our case, the requirement of 
orbifold symmetry allows only one of these solutions to be present. 

The solutions of the differential equations (\ref{eigen1}) for the 
case $x_n>0$ are Bessel functions. For convenience we assume that 
$\nu\ne\frac12+N$ with an integer $N$. Then the most general 
solutions can be written in the form
\begin{equation}
   f_n^{L,R}(t) = \sqrt{t} \left[ a_n^{L,R}\,J_{\frac12\mp\nu}(x_n t)
   + b_n^{L,R}\,J_{-\frac12\pm\nu}(x_n t) \right] \,.
\end{equation}
For the special values $\nu=\frac12+N$ the solutions are 
superpositions of Bessel functions of the first and second kind, 
which can obtained from our results using a limiting procedure. The 
two functions $f_n^L(t)$ and $f_n^R(t)$ are not independent, since 
they are coupled by the first-order differential equations in 
(\ref{cond1}), which imply $b_n^L=a_n^R$ and $b_n^R=-a_n^L$. Hence, 
the solutions take the form 
\begin{eqnarray}\label{fLfR}
   f_n^L(t) &=& \sqrt{t} \left[ a_n^L\,J_{\frac12-\nu}(x_n t)
    + a_n^R\,J_{-\frac12+\nu}(x_n t) \right] \,, \nonumber\\
   f_n^R(t) &=& \sqrt{t} \left[ a_n^R\,J_{\frac12+\nu}(x_n t)
    - a_n^L\,J_{-\frac12-\nu}(x_n t) \right] \,.
\end{eqnarray}

\TABULAR[t]{|l|c|cc|c|}
{\hline
Option & Eigenvalues $x_n>0$ & $a_n^L$ & $a_n^R$ & Zero Mode \\
\hline
L, $\nu<\frac12$ & $J_{\frac12-\nu}(x_n)=0$
 & ${\cal N}_{\frac32-\nu}(x_n)$ & 0 & R \\
L, $\nu>\frac12$ & $J_{\nu-\frac12}(x_n)=0$ & 0 
 & ${\cal N}_{\frac12+\nu}(x_n)$ & R \\
R & $J_{\frac12+\nu}(x_n)=0$ & 0 & ${\cal N}_{\frac32+\nu}(x_n)$
 & L \\
\hline}
{\label{tab:1}
Bulk fermion solutions for the two choices of boundary conditions, 
in the limit where $\epsilon=e^{-k r_c\pi}\to 0$. The wave functions 
$f_n^{L,R}(t)$ take the form (\protect\ref{fLfR}) with coefficients 
$a_n^{L,R}$ given in the third and fourth columns.}

To proceed we must specify the boundary conditions at the locations 
of the 3-branes. This will give rise to a discrete spectrum of 
Kaluza--Klein modes. Orbifold symmetry allows two choices of boundary 
conditions: either all left-handed fields are odd under $\phi$-parity 
and all right-handed ones even (``option L''), or all right-handed 
fields are odd and all left-handed ones even (``option R''). In the
first case the boundary conditions are $f_n^L(\epsilon)=f_n^L(1)=0$, 
and in the second case $f_n^R(\epsilon)=f_n^R(1)=0$. Which of these 
choices is realized in nature is a question that cannot be answered 
without understanding the physics on the 3-branes, which is beyond 
the scope of the field-theory model suggested in \cite{RS1}. The two 
cases are straightforward to analyze. Using the asymptotic behavior 
of the Bessel functions, $J_n(x)\sim x^n$ as $x\to 0$, it follows 
that in the limit $\epsilon\to 0$ only one of the two terms in the 
wave functions in (\ref{fLfR}) remains. Taking $\epsilon\to 0$ is an 
excellent approximation unless we were to consider integrals of the 
functions $f_n^{L,R}(t)$ with weight functions that are singular as 
$t\to 0$. Table~\ref{tab:1} summarizes the results for the eigenvalues 
and eigenfunctions of the non-zero modes in that limit. The solutions 
shown correspond to positive $\phi$ and must be extended to negative 
$\phi$ in accordance with the orbifold symmetry. For option L the 
results take a different form depending on whether $\nu<\frac12$ or 
$\nu>\frac12$, as indicated. The second column in the table shows the 
equation that determines the eigenvalues $x_n$. In the next two 
columns we give the values of the coefficients $a_n^{L,R}$ of the 
properly normalized solutions. The normalization constants 
${\cal N}_a(x)$ obey $|{\cal N}_a(x)|^2=2/[J_a(x)]^2$. The last column 
shows the chirality of the zero mode. The zero-mode wave functions 
can be recovered by taking the limit $x_n\to 0$; however, the 
normalization constants do not apply in this case. In 
figure~\ref{fig:1} we show the first few Kaluza--Klein modes for 
option L and two values of the parameter $\nu$ just below or above the 
critical value $\nu=\frac12$. The important point to notice is the 
localization of the right-handed zero mode $f_0^R(t)$ on the hidden 
brane (at $t=\epsilon$) for $\nu>\frac12$.

\FIGURE[t]{\epsfig{file=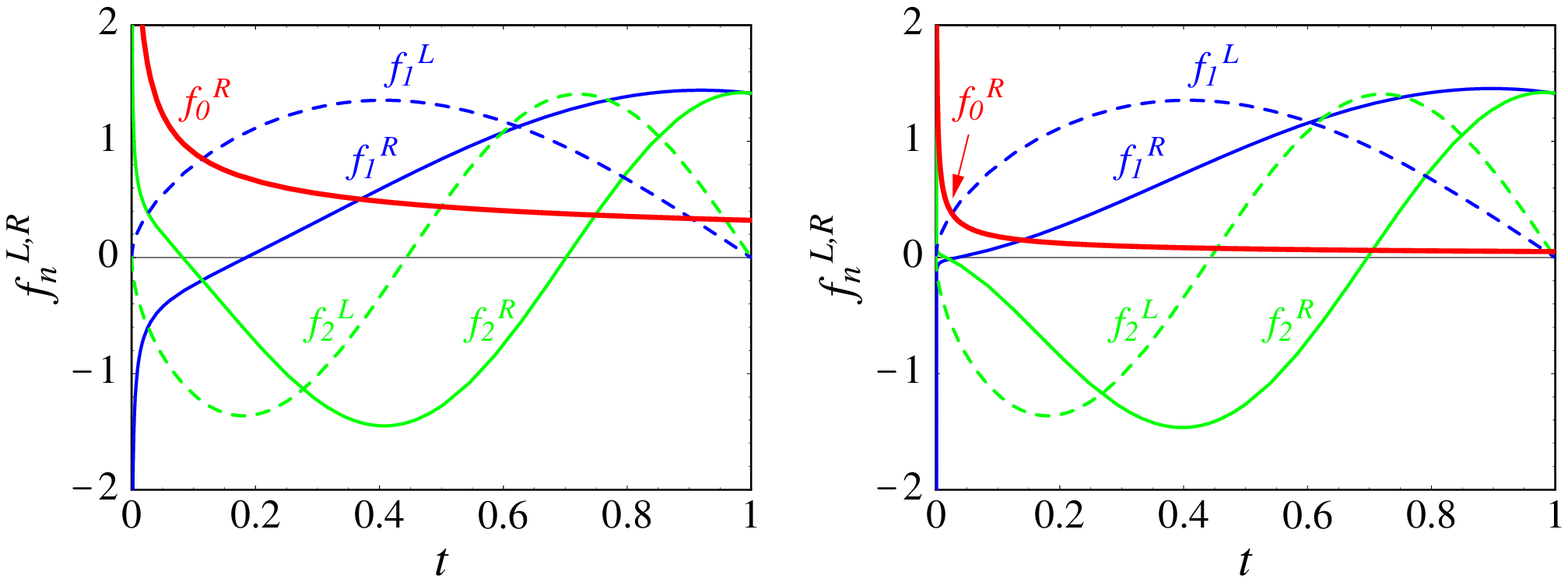,width=14cm}
\caption{\label{fig:1}
Right-handed (solid) and left-handed (dashed) Kaluza--Klein modes with 
$n\le 2$ for $\nu=m/k=0.45$ (left) and 0.55 (right), with boundary 
conditions such that all left-handed fields vanish at $t=\epsilon$ and
$t=1$ (option L). We show exact results obtained with 
$\epsilon=10^{-16}$. In both cases, the corresponding eigenvalues are 
$x_1\approx 2.49$ and $x_2\approx 5.60$.}}

For the special case of integer $\nu$ the exact solutions for the wave 
functions can be expressed in terms of trigonometric functions. As an 
example, we quote results for $\nu=0$ and $\nu=1$ choosing for the 
boundary conditions option L, which will be of special relevance to 
our discussion below. In both cases the non-zero eigenvalues are given 
by $x_n=n\pi/(1-\epsilon)$ with an integer $n\ge 1$, and the 
left-handed solutions are $f_n^L(t)={\cal N}\sin[x_n(t-\epsilon)]$, 
where $|{\cal N}|^2=2/(1-\epsilon)$. For $\nu=0$, the right-handed 
solutions are given by
\begin{equation}
   f_0^R(t) = \frac{\cal N}{\sqrt2} \,, \qquad
   f_n^R(t) = - {\cal N}\cos[x_n(t-\epsilon)] \,,
\end{equation}
whereas for $\nu=1$ they take the form
\begin{equation}
   f_0^R(t) = \frac{\cal N}{\sqrt2}\,\frac{\sqrt\epsilon}{t} \,,
   \qquad
   f_n^R(\tau) = {\cal N} \left( \frac{\sin[x_n(t-\epsilon)]}{x_n t}
   - \cos[x_n(t-\epsilon)] \right) \,.
\end{equation}

\section{Yukawa interactions and neutrino phenomenology}

We will now show how including a sterile bulk fermion in the RS model 
can provide a mechanism for obtaining small Dirac neutrino masses, 
which is quite different from the see-saw mechanism. We focus first on 
a single fermion generation and consider a scenario where all matter 
and gauge fields charged under the Standard Model gauge group are 
confined to the visible brane at $\phi=\pi$, whereas a gauge-singlet 
fermion field propagates in the bulk. After integration over the 
compact extra dimension we obtain a tower of four-dimensional 
Kaluza--Klein fermions in the four-dimensional theory, as shown in 
(\ref{Sferm}). We choose boundary conditions such that there is 
a right-handed zero mode (option L) with wave function $f_0^R(t)$ 
given in (\ref{zero}). Only this choice will lead to an interesting
neutrino phenomenology.

Omitting gauge interactions, the action for a Higgs doublet 
$H=(\phi_1,\phi_2)$, a left-handed lepton doublet $L=(\nu_L,e_L)$ 
and a right-handed lepton $e_R$ localized on the visible brane is 
\begin{eqnarray}
   S &=& \int\!\mbox{d}^4x\,\sqrt{-g_{\rm vis}} \left\{
    g_{\rm vis}^{\mu\nu}\,\partial_\mu H_0^\dagger\,\partial_\nu H_0
    - \lambda \left( |H_0|^2 - v_0^2 \right)^2 \right\} \nonumber\\
   &+& \int\!\mbox{d}^4x\,\sqrt{-g_{\rm vis}} \left\{
    \bar L_0\hat\gamma^\mu\partial_\mu L_0
    + \bar e_{R0}\hat\gamma^\mu\partial_\mu e_{R0}
    - \left( y_e \bar L_0 H_0 e_{R0} + \mbox{h.c.} \right)
    \right\} \,,
\end{eqnarray}
where $g_{\rm vis}^{\mu\nu}=e^{2k r_c\pi}\,\eta^{\mu\nu}$ is the 
induced metric on the brane, $\sqrt{-g_{\rm vis}}
=\mbox{det}(-g_{\mu\nu}^{\rm vis})=e^{-4k r_c\pi}$, and 
$\hat\gamma^\mu=E_a^\mu(\phi=\pi)\,\gamma^a=e^{k r_c\pi}\,\gamma^\mu$.
To restore a canonical normalization of the fields on the brane we 
must perform the rescalings $H_0\to e^{k r_c\pi}\,H$, $L_0\to 
e^{\frac32 k r_c\pi}\,L$ and $e_{R0}\to e^{\frac32 k r_c\pi}\,e_R$, 
upon which the action takes the form
\begin{eqnarray}
   S &=& \int\!\mbox{d}^4x \left\{
    \partial_\mu H^\dagger \partial^\mu H
    - \lambda \left( |H|^2 - v^2 \right)^2 \right\} \nonumber\\
   &+& \int\!\mbox{d}^4x \left\{ \bar L\,i\rlap/\partial\,L
    + \bar e_R\,i\rlap/\partial\,e_R - \left( y_e \bar L H e_R 
    + \mbox{h.c.} \right) \right\} \,,
\end{eqnarray}
where $v=e^{-k r_c\pi}\,v_0$. The remarkable feature noted in 
\cite{RS1} is that all dimensionful parameters such as the Higgs 
vacuum expectation value get rescaled by the warp factor and turned 
from Planck-scale into weak-scale couplings, whereas dimensionless 
parameters such as $\lambda$ and $y_e$ remain unchanged. 

We now introduce a Yukawa coupling of the bulk fermion with the Higgs 
and lepton fields. With our choice of boundary conditions all 
left-handed Kaluza--Klein modes vanish at the visible brane, so only 
the right-handed modes can couple to the Standard Model fields on the 
brane. However, in a more realistic scenario which takes into account 
a finite width of the 3-branes there will most likely be a non-zero 
(and indeed sizeable) overlap of the left-handed modes with the 
Standard Model fields. Hence, in order to avoid weak-scale neutrino
masses and lepton-number violating interactions we assign lepton 
number $L=1$ to the bulk fermion state. Then the only gauge-invariant 
coupling is of the form
\begin{equation}
   S_Y = - \int\!\mbox{d}^4x\,\sqrt{-g_{\rm vis}} \left\{
   \hat Y_5 \bar L_0(x) \widetilde H_0(x) \Psi_R(x,\pi)
   + \mbox{h.c.} \right\} \,,
\end{equation}
where $\widetilde H=i\sigma_2 H^*$, and the Yukawa coupling $\hat Y_5$ 
is naturally of order $M^{-1/2}$, with $M$ the fundamental Planck 
scale of the theory. Rescaling the Standard Model fields in the way 
described above, and inserting for the bulk fermion the Kaluza--Klein 
ansatz (\ref{KK}), we find
\begin{equation}
   S_Y = - \sum_{n\ge 0} \int\!\mbox{d}^4x \left\{ y_n 
   \bar L(x) \widetilde H(x) \psi_n^R(x) + \mbox{h.c.} \right\}
\end{equation}
with the effective Yukawa couplings
\begin{equation}\label{yeff}
   y_n =\sqrt{k}\,\hat Y_5\,f_n^R(1) \equiv Y_5\,f_n^R(1) \,,
\end{equation}
where $Y_5$ is naturally of order unity. After electroweak symmetry 
breaking, this Yukawa interaction gives rise to a neutrino mass term 
$\bar\psi_L^\nu\,{\cal M}\,\psi_R^\nu + \mbox{h.c.}$, which in the 
basis $\psi_L^\nu=(\nu_L,\psi_1^L,\dots,\psi_n^L)$ and 
$\psi_R^\nu=(\psi_0^R,\psi_1^R,\dots,\psi_n^R)$, with $n\to\infty$, 
takes the form
\begin{equation}
   {\cal M} = \pmatrix{
    v y_0 & v y_1 & \dots & v y_n \cr
    0 & m_1 & \dots & 0 \cr
    \vdots & 0 & \ddots & 0 \cr
    0 & 0 & \dots & m_n} \,.
\end{equation}
As a consequence, there will be a mixing of the Standard Model
neutrino $\nu_L$ with the heavy, sterile (with respect to the 
Standard Model gauge interactions) bulk neutrinos $\psi_n^L$. The 
Kaluza--Klein excitations of the bulk fermion have masses $m_n$ of 
order the weak scale $v$. Thus, in order to obtain a light neutrino we 
need $|y_0|\ll 1$, which requires having a very small wave function 
of the zero mode on the visible brane, i.e., $|f_0^R(1)|\ll 1$. But 
this is precisely what happens if the fundamental fermion mass $m$ 
satisfies the condition $m>k/2$. Since, as mentioned earlier, the 
curvature $k$ must be smaller than the fundamental scale $M$, this is 
a natural requirement in the context of the RS model. 

In order to study the properties of the physical neutrino states we 
diagonalize the squared mass matrix ${\cal M M}^\dagger$. The 
eigenvalues of this matrix are the squares of the physical neutrino 
masses, and the unitary matrix $U$ defined such that 
$U^\dagger {\cal M M}^\dagger U$ is diagonal determines the 
left-handed neutrino mass eigenstates via $\psi_L^\nu
=U\psi_L^{\rm phys}$. We denote by $m_\nu$ the mass of the lightest 
neutrino $\nu_L^{\rm phys}$ and define a mixing angle $\theta_\nu$ 
such that $\nu_L=\cos\theta_\nu\,\nu_L^{\rm phys}+\dots$, where the 
dots represent the admixture of heavy, sterile bulk states. To leading 
order in the small parameter $|y_0|\ll 1$ we obtain
\begin{equation}
   m_\nu = v |y_0|\cos\theta_\nu \,, \qquad
   \tan^2\!\theta_\nu = \sum_{n\ge 1} \frac{v^2 |y_n|^2}{m_n^2} \,.
\end{equation}
Since the mixing angle is constrained by experiment to be very small
(see below), it follows from (\ref{zero}) and (\ref{yeff}) that
\begin{equation}
   m_\nu\simeq \sqrt{2\nu-1}\,|Y_5|\,\epsilon^{\nu-\frac12}\,v
   \sim M \left( \frac{v}{M} \right)^{\nu+\frac12} \,; \quad 
   \nu>\frac12 \,.
\end{equation}
This result is remarkable, as it provides a parametric dependence
of the neutrino mass on the ratio of the electroweak and Planck  
scales that is different from the see-saw relation $m_\nu\sim v^2/M$, 
except for the special case where $\nu=\frac32$. This flexibility
allows us to reproduce a wide range of neutrino masses without any
fine tuning. For instance, taking $v/M=10^{-16}$, the 
phenomenologically interesting range of $m_\nu$ between $10^{-5}$\,eV 
and 10\,eV can be covered by varying $\nu$ between 1.1 and 1.5. 

The measurement of the invisible width of the $Z^0$ boson, which 
yields $n_\nu=2.985\pm 0.008$ for the apparent number of light 
neutrinos \cite{Sirl}, implies that the mixing angle $\theta_\nu$ 
must be of order a few percent. For instance, assuming an equal 
admixture of sterile neutrinos for the three generations of light 
neutrinos, we obtain $n_\nu=3\cos^2\!\theta_\nu$ and hence 
$\tan^2\!\theta_\nu=0.005\pm 0.003$. From table~\ref{tab:1} it 
follows that with our choice of boundary conditions the wave 
functions of all excited right-handed Kaluza--Klein modes obey 
$|f_n^R(1)|=\sqrt{2}$ (for $\epsilon\to 0$). We thus obtain
\begin{equation}
   \tan^2\!\theta_\nu = \frac{v^2 |Y_5|^2}{(\epsilon k)^2}
   \sum_{n=1}^\infty \frac{2}{x_n^2}
   = \frac{1}{2\nu+1}\,\frac{v_0^2 |Y_5|^2}{k^2} \,,
\end{equation}
where $x_n$ are the roots of $J_{\nu-\frac12}(x_n)=0$. The infinite 
sum can be performed exactly and yields $1/(2\nu+1)$. To satisfy the 
bound on the mixing angle for $\nu=O(1)$ requires that 
$v_0|Y_5|/k\lsim 0.1$, which is possible without much fine tuning. We 
emphasize, however, that it would be unnatural to have the 
dimensionless combination $v_0|Y_5|/k$ much less than unity, so a 
mixing angle $\theta_\nu$ not much smaller than the current 
experimental bound is a generic feature of our scenario, which can be 
tested by future precision measurements.

So far we have shown how a right-handed bulk fermion can give a small 
Dirac mass to a Standard Model neutrino. We now generalize this 
mechanism to three neutrino flavors and more than one bulk fermion. 
Interestingly, such a generalization is forced upon us by the 
requirement that the parity anomaly for fermions in an odd number of
dimensions vanish. When an odd number of bulk fermions in five 
dimensions are coupled to a gauge field or gravity, the $\phi$-parity 
of the action (\ref{action}) is broken at the quantum level 
\cite{Redl,Call}. To obtain a minimal model that is anomaly free we 
thus introduce two bulk fermions with boundary conditions option L, 
so there are two massless right-handed zero modes.\footnote{More 
complicated models with four or more bulk states are possible. These 
states could be subject to different boundary conditions. If we impose 
lepton number, only the right-handed modes can couple to the Standard 
Model fields. At least two right-handed zero modes are needed for a 
successful neutrino phenomenology.} 
In order to explain the atmospheric and solar neutrino anomalies in 
terms of neutrino oscillations one needs two very different 
mass-squared differences: $\Delta m_{21}^2\ll\Delta m_{32}^2$, where 
$\Delta m_{ij}=m_{\nu_i}^2-m_{\nu_j}^2$, and by convention 
$m_{\nu_1}<m_{\nu_2}<m_{\nu_3}$. This requires a minimum of two 
massive neutrinos; however, the third neutrino can be massless. In 
our minimal model this is indeed what happens. Although it is perhaps 
unconventional to consider a scenario where the number of 
right-handed neutrinos does not match the number of left-handed ones, 
we will see that our model explains successfully the known features 
of the neutrino mass and mixing parameters.

In order to explore this minimal model in more detail we ignore, for 
simplicity, the heavy Kaluza--Klein excitations of the bulk fermions
and focus only on the zero modes. As mentioned above, the admixture 
of weak-scale sterile neutrino states must be strongly suppressed. It
is natural to allow for the possibility that the two bulk fermions 
have different masses $m_1>m_2$ (of order the Planck scale) in the 
fundamental theory, and that they couple with similar strength to the 
three left-handed neutrino flavors. According to (\ref{zero}) and 
(\ref{yeff}), the effective Yukawa couplings of the two right-handed 
zero modes $\psi_0^{R,1}$ and $\psi_0^{R,2}$ can then be parameterized 
as $x_f\,\epsilon^{\nu_1-\frac12}$ and $y_f\,\epsilon^{\nu_2-\frac12}$ 
with $\nu_i=m_i/k$ (for $i=1,2$) and flavor-dependent couplings $x_f$, 
$y_f$ (with $f=e,\mu,\tau$) of order unity. Note that the Yukawa 
couplings of the two zero modes have a very different magnitude: 
$x_f/y_f=O(\epsilon^{\nu_1-\nu_2})$. The resulting neutrino mass term 
$\bar\psi_L^\nu\,{\cal M}\,\psi_R^\nu+\mbox{h.c.}$ in the truncated 
basis $\psi_L^\nu=(\nu_e^L,\nu_\mu^L,\nu_\tau^L)$ and 
$\psi_R^\nu=(\psi_0^{R,1},\psi_0^{R,2})$ is
\begin{equation}
   {\cal M} = v\,\epsilon^{\nu_2-\frac12} \pmatrix{
    \epsilon^{\nu_1-\nu_2}\,x_e & y_e \cr
    \epsilon^{\nu_1-\nu_2}\,x_\mu & y_\mu \cr
    \epsilon^{\nu_1-\nu_2}\,x_\tau & y_\tau} \,.
\end{equation}
Diagonalizing the matrix ${\cal M M^\dagger}$ to leading order in 
$\epsilon$ we find that the physical neutrino mass eigenstates 
comprise a massless left-handed neutrino $\nu_1$, a very light Dirac 
neutrino with mass squared
\begin{equation}\label{m2sq}
   m_{\nu_2}^2 = v^2\,\epsilon^{2\nu_1-1}\,
   \frac{|[e\mu]|^2+|[\mu\tau]|^2+|[\tau e]|^2}
        {|y_e|^2+|y_\mu|^2+|y_\tau|^2}
   \sim M^2 \left( \frac{v}{M} \right)^{2\nu_1+1} \,,
\end{equation}
and a light Dirac neutrino with mass squared
\begin{equation}
   m_{\nu_3}^2 = v^2\,\epsilon^{2\nu_2-1}
   \left( |y_e|^2+|y_\mu|^2+|y_\tau|^2 \right)
   \sim M^2 \left( \frac{v}{M} \right)^{2\nu_2+1} \,.
\end{equation}
In (\ref{m2sq}) we use the short-hand notation 
$[ij]\equiv x_i y_j-x_j y_i$. Since the lightest neutrino is massless
it follows that $\Delta m_{21}^2=m_{\nu_2}^2$ and 
$\Delta m_{32}^2\simeq m_{\nu_3}^2$, and the ratio 
$\Delta m_{21}^2/\Delta m_{32}^2\sim (v/M)^{2(\nu_1-\nu_2)}$. An 
interpretation of the solar neutrino anomaly in terms of neutrino 
oscillations based on the MSW effect \cite{MSW} yields values of 
$\Delta m_{21}^2$ in the range $10^{-6}$--$10^{-5}$\,eV$^2$, whereas 
oscillations in vacuum would require a smaller value of order 
$10^{-10}$\,eV$^2$ \cite{SN}. Such masses can be reproduced in our 
model by setting $\nu_1\approx 1.34$--1.37 and $\nu_1\approx 1.5$, 
respectively. An explanation of the atmospheric neutrino anomaly in 
terms of neutrino oscillations yields $\Delta m_{32}^2$ in the range 
$5\cdot 10^{-4}$--$6\cdot 10^{-3}$\,eV$^2$ \cite{SK}, which we can 
reproduce by taking $\nu_2\approx 1.27$--1.29. In other words, we 
can understand the observed hierarchy of the experimentally favored 
neutrino masses in terms of a small difference of the bulk fermion
masses in the fundamental theory. Note that in our minimal model
the lightest neutrino is massless. This can be changed by introducing 
four (or more) bulk fermion states with more than two right-handed 
zero modes, in which case also the lightest neutrino becomes massive,
with $m_{\nu_1}^2\ll m_{\nu_2}^2$.

Despite the fact that a strong neutrino mass hierarchy is a generic
feature of our model, the mixing matrix $U$ relating the neutrino 
flavor and mass eigenstates does not contain any small parameter. 
Defining $\nu_f=\sum_{i=1}^3 U_{fi}\,\nu_i$ we find that all the 
entries $U_{fi}$ are of order unity. In the limit $\epsilon\to 0$ we 
obtain
\begin{equation}
   U = \pmatrix{
    U_{e1} & U_{e2} & U_{e3} \cr
    U_{\mu 1} & U_{\mu 2} & U_{\mu 3} \cr
    U_{\tau 1} & U_{\tau 2} & U_{\tau 3}}
   = \pmatrix{
    \frac{[\mu\tau]^*}{N_1} &
    \frac{y_\mu^* [e\mu]-y_\tau^* [\tau e]}{N_1 N_2} &
    \frac{y_e}{N_2} \cr
    \frac{[\tau e]^*}{N_1} &
    \frac{y_\tau^* [\mu\tau]-y_e^* [e\mu]}{N_1 N_2} &
    \frac{y_\mu}{N_2} \cr
    \frac{[e\mu]^*}{N_1} &
    \frac{y_e^* [\tau e]-y_\mu^* [\mu\tau]}{N_1 N_2} & 
    \frac{y_\tau}{N_2}} \,,
\end{equation}
where $N_1^2=|[e\mu]|^2+|[\mu\tau]|^2+|[\tau e]|^2$ and
$N_2^2=|y_e|^2+|y_\mu|^2+|y_\tau|^2$. A mixing matrix of this type,
which lacks the strong hierarchy of the quark mixing matrix, can 
account for the experimental constraints on the neutrino mixing 
angles. (In fact, it has been pointed out that a fair fraction of 
random Dirac mixing matrices is consistent with these constraints 
\cite{Hito}.) The precise form of these constraints depends on how
the data are analyzed, and whether the solar and atmospheric 
neutrino anomalies individually are interpreted in terms of 
two-neutrino or three-neutrino mixing. Constraints from the CHOOZ 
reactor experiment \cite{CHOOZ} combined with the atmospheric 
neutrino data imply that $|U_{e3}|^2\lsim\mbox{few \%}$ \cite{LLMS}, 
which means that $|y_e|$ should be less than $|y_\mu|$ and 
$|y_\tau|$. In the limit where $|y_e|^2\ll|y_\mu|^2+|y_\tau|^2$, the 
mixing angles $\theta_{12}$ and $\theta_{23}$ responsible for 
$\nu_e\leftrightarrow\nu_\mu$ and $\nu_\mu\leftrightarrow\nu_\tau$ 
oscillations obey the approximate relations
\begin{eqnarray}
   \sin^2\!2\theta_{12}
   &\simeq& \frac{4|x_e|^2 (|y_\mu|^2+|y_\tau|^2)\,|[\mu\tau]|^2}
             {\left[ |x_e|^2 (|y_\mu|^2+|y_\tau|^2) + |[\mu\tau]|^2
              \right]^2} \,, \nonumber\\
   \sin^2\!2\theta_{23} &\simeq&
    \frac{4|y_\mu|^2 |y_\tau|^2}{\left(|y_\mu|^2+|y_\tau|^2\right)^2}
    \,.
\end{eqnarray} 
The atmospheric neutrino anomaly is best explained by near-maximal 
$\nu_\mu\leftrightarrow\nu_\tau$ mixing, such that 
$\sin^2\!2\theta_{23}>0.82$ \cite{AN3}. This implies 
$0.64<|y_\mu/y_\tau|<1.57$, which clearly does not pose a problem 
for our model. Likewise, a large-mixing-angle solution to the 
solar neutrino problem requires $\sin^2\!2\theta_{12}>0.75$ \cite{SN}, 
which yields $0.58<|x_e|\sqrt{|y_\mu|^2+|y_\tau|^2}/|[\mu\tau]|<1.73$. 
The small-mixing-angle MSW solution, on the other hand, prefers
$\sin^2\!2\theta_{12}\sim 10^{-2}$, which would require that the 
quantities $|x_e|\sqrt{|y_\mu|^2+|y_\tau|^2}$ and 
$|x_\mu y_\tau-x_\tau y_\mu|$ differ by about a factor 20. This could
either be achieved by having $|x_e|\ll|x_{\mu,\tau}|$, or via a near 
degeneracy of $x_\mu y_\tau$ and $x_\tau y_\mu$.

\section{Conclusions}

We have studied bulk fermion solutions in the localized gravity 
model with non-factorizable geometry introduced by Randall and 
Sundrum to solve the gauge-hierar\-chy problem. Similar to the case of 
scalar or vector fields propagating in the extra compact dimension, 
we have found that the Kaluza--Klein modes have weak-scale masses even 
if the fermion mass in the fundamental, five-dimensional theory is of 
order the Planck scale. However, a distinct feature of bulk fermion 
solutions is the possible presence of zero modes due to the fact 
that the 3-branes in the Randall--Sundrum model act as domain walls. 

Our most important finding is that, if the fundamental mass $m$ in the 
five-dimensional theory is larger than half the curvature $k$ of the 
compact space, an appropriate choice of the orbifold boundary 
conditions leads to a right-handed zero mode localized on the hidden 
brane, whose wave function on the visible brane is power-suppressed in 
the ratio of the weak scale to the fundamental Planck scale. Coupling 
the Higgs and left-handed lepton fields of the Standard Model, 
localized on the visible brane, with a bulk right-handed neutrino 
provides a new mechanism for obtaining small neutrino masses. 
Remarkably, this mechanism leads to a generalization of the see-saw 
formula with a different parametric dependence on the ratio $v/M$, 
which can easily reproduce neutrino masses in the range $10^{-5}$\,eV 
to 10\,eV. Without much fine tuning the mixing of the Standard Model 
left-handed neutrino with sterile, weak-scale Kaluza--Klein 
excitations of the bulk fermion can be made consistent with 
experimental bounds. However, a generic prediction of our model is 
that such a mixing should occur at a level not much below the present 
bound.

Finally, we have shown that with an even number of bulk fermions one
can obtain viable models of neutrino flavor oscillations, which 
naturally predict a mass hierarchy and a neutrino mixing matrix not 
containing any small parameter. A minimal implementation of this 
scenario consists of two right-handed neutrinos, identified with the 
zero modes of two bulk fermions with slightly different masses in the 
five-dimensional theory, coupled to the three left-handed neutrinos of 
the Standard Model. In this model we obtain a massless left-handed 
neutrino and two massive Dirac neutrinos with a large mass hierarchy 
and generically large mixing angles.

\acknowledgments
We are grateful to M.~Schmaltz, H.~Davoudiasl, L.~Dixon, Y.~Nir, 
M.~Peskin and T.~Rizzo for useful discussions. Y.G.~is supported by 
the Department of Energy under contract DE--AC03--76SF00515. M.N.~is 
supported in part by the National Science Foundation.

\end{document}